\documentclass[pre,amsmath,superscriptaddress,nofootinbib,preprintnumbers]{revtex4}

\usepackage{graphicx}
\usepackage{bm}

\begin{document}

\title{Rare events and the convergence of exponentially averaged work values}

\author{Christopher Jarzynski}

\affiliation{Theoretical Division, T-13, MS B213, Los Alamos National Laboratory,
Los Alamos, New Mexico 87545 \\
{\tt chrisj@lanl.gov}
}

\preprint{LAUR-06-004}

\begin{abstract}
Equilibrium free energy differences are given by exponential averages
of nonequilibrium work values;
such averages, however, often converge poorly, as they are dominated
by rare realizations.
I show that there is a simple and intuitively appealing description of these
rare but dominant realizations.
This description is expressed as a duality between ``forward''
and ``reverse'' processes, and
provides both heuristic insights and quantitative estimates regarding
the number of realizations needed for
convergence of the exponential average.
Analogous results apply to the equilibrium
perturbation method of estimating free energy differences.
The pedagogical example of a piston and gas
[R.C.~Lua and A.Y.~Grosberg, J.\ Phys.\ Chem.\ B {\bf 109}, 6805 (2005)]
is used to illustrate the general discussion.
\end{abstract}

\maketitle

The nonequilibrium work theorem,
\begin{equation}
\label{eq:nwt0}
\Bigl\langle e^{-\beta W}\Bigr\rangle = e^{-\beta\Delta F},
\end{equation}
relates the work performed on a system during a nonequilibrium process,
to the free energy difference between two equilibrium states of that system.
The angular brackets denote an average over an ensemble of realizations (repetitions)
of a thermodynamic process,
during which a system evolves in time as a control parameter $\lambda$
is varied from an initial value $A$ to a final value $B$.
$W$ is the external work performed on the system during one realization;
$\Delta F = F_B-F_A$ is the free energy difference between two equilibrium
states of the system, corresponding to $\lambda=A$ and $B$;
and $\beta$ is the inverse temperature of a heat reservoir
with which the system is equilibrated prior to the start of each realization of
the process.
A sample of derivations of Eq.~\ref{eq:nwt0} can be found in
Refs.~\cite{Jarzynski97,Crooks98,Crooks99,HummerSzabo01,Neal01,Sun03,Evans03,Oberhofer05,Jarzynski04,ImparatoPeliti05};
pedagogical and review treatments are given in
Refs.~\cite{Jarzynski.Ladek.02,frenkelSmit,Ritort03,ParkSchulten04,HummerSzabo05,BustamanteLiphardtRitort05};
for experimental tests of this and closely related results, see
Refs.~\cite{Liphardt_etal02,DouarcheCilibertoPetrosyanRabbiosi05,DouarcheCilibertoPetrosyan05,Collin05};
finally, Refs.~\cite{Yukawa00,Kurchan00,Mukamel03,deRoeck04,Monnai05}
discuss quantal versions of the theorem.

In principle, Eq.\ref{eq:nwt0} implies that $\Delta F$
can be estimated using nonequilibrim experiments or numerical simulations.
If we repeat the thermodynamic process $N$ times, and observe work values
$W_1,\,W_2,\cdots,\,W_N$, then
\begin{equation}
\label{eq:estDF}
\Delta F \approx -\beta^{-1}\ln
\Bigl[
\frac{1}{N} \sum_{n=1}^N
e^{-\beta W_n}
\Bigr],
\end{equation}
where the approximation becomes an equality in the limit of infinitely
many realizations, $N\rightarrow\infty$.
In practice, the average of $e^{-\beta W}$ is often dominated by very rare realizations,
leading to poor convergence with $N$.
The aim of this paper is develop an understanding of these
rare but important realizations.
I will argue that there is a simple description 
of these dominant realizations, which leads to both quantitative estimates
and useful heuristic insights regarding the number of realizations needed for
convergence of the average of $e^{-\beta W}$.

The organization of this paper is as follows.
In Section \ref{sec:summary} the central result is summarized, then illustrated
using a simple example.
Section \ref{sec:derivation} contains a derivation of this result.
Section \ref{sec:numbers} discusses the number of realizations needed for
convergence of the exponential average.
Section \ref{sec:fep} focuses on the free energy perturbation method
(a limiting case of Eq.\ref{eq:nwt0}),
and Section \ref{sec:discussion} concludes with a brief discussion.

\section{Summary and illustration of central result}
\label{sec:summary}

\begin{figure}[htbp]
  \centering
  \includegraphics[scale=0.45,angle=-90]{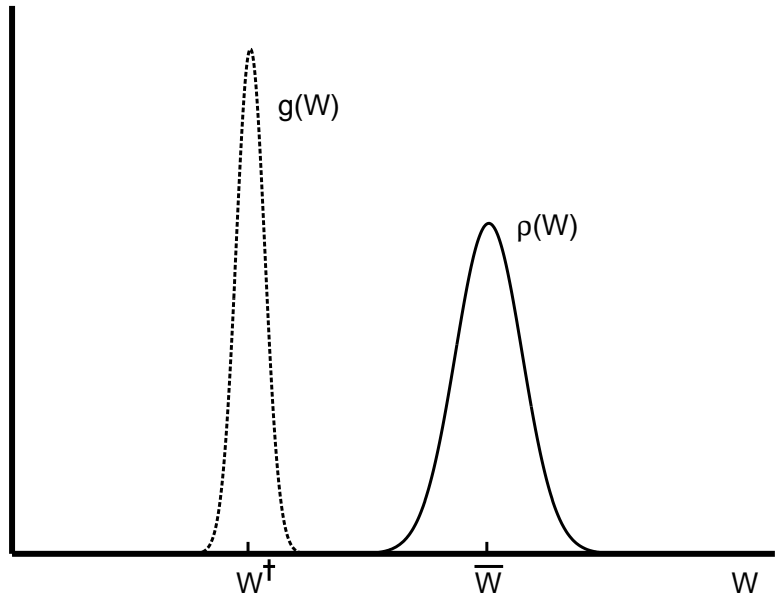}
  \caption{
During most realizations of the process, we observe work values near the
peak of $\rho(W)$.
However, the average of $e^{-\beta W}$ is dominated by
realizations for which the work is observed to be in the region
around the peak of $g(W) = \rho(W) e^{-\beta W}$.
}
  \label{fig:rho.and.g}
\end{figure}

The average in Eq.\ref{eq:nwt0} can be written as
\begin{equation}
\label{eq:nwt_int}
\Bigl\langle e^{-\beta W}\Bigr\rangle =
\int dW\,\rho(W)\,e^{-\beta W} \equiv \int dW\,g(W) ,
\end{equation}
where $\rho(W)$ is the ensemble distribution of work values.
Fig.\ref{fig:rho.and.g} shows a schematic plot of this distribution,
and of the integrand in Eq.\ref{eq:nwt_int},
$g(W) = \rho(W) \,e^{-\beta W}$.
While $\rho$ is peaked near the mean of the distribution,
\begin{equation}
\label{eq:wbar}
\overline{W} = \int dW\,\rho(W)\,W,
\end{equation}
$g$ is peaked around a lower value, $W^\dagger$,
as the factor $e^{-\beta W}$ has the effect of strongly weighting
those work values that are in the far left tail of $\rho$.
Explicitly,
\begin{equation}
\label{eq:wdag}
W^\dagger = c^{-1}\,\int dW\,g(W)\,W ,
\end{equation}
where $c = \int dW\,g(W)$.
As pointed out by Ritort in the context of a ``trajectory thermodynamics''
formalism~\cite{Ritort04},
the average of $e^{-\beta W}$ is dominated by the region near the peak of the integrand
of Eq.\ref{eq:nwt_int}, i.e.\ near $W^\dagger$.
I will use the term {\it typical} to refer to those
realizations whose work values are near the peak of $\rho$,
and {\it dominant} to refer to those for which
the work is near the peak of $g$.
Typical realizations are the ones that we ordinarily
observe when carrying out the process,
while dominant realizations are those rare realizations
that contribute the greatest share to the average of $e^{-\beta W}$.

For the process discussed in the previous paragraph, the work parameter
$\lambda$ is varied from $A$ to $B$.
Following Crooks~\cite{Crooks98}, let us also consider a process during which $\lambda$ is
manipulated from $B$ to $A$,
and let us use the terms {\it forward} (F) and {\it reverse} (R) 
to distinguish between the two processes.
Specifically, if $\lambda_t^F$ denotes the schedule for varying the work parameter
during the forward process, from $\lambda_0^F=A$ to $\lambda_\tau^F=B$,
then the reverse process is defined by the schedule
\begin{equation}
\label{eq:for_rev_schedules}
\lambda_t^R = \lambda_{\tau-t}^F,
\end{equation}
where $\tau$ is the duration of either process.

Note that the nonequilibrium work theorem applies to both processes:
\begin{equation}
\label{eq:both}
\Bigl\langle e^{-\beta W}\Bigr\rangle_F = e^{-\beta\Delta F} \qquad,\qquad
\Bigl\langle e^{-\beta W}\Bigr\rangle_R = e^{+\beta\Delta F} ,
\end{equation}
where $\langle\cdots\rangle_{F/R}$ denotes an average over realizations of the
forward/reverse process;
and by convention $\Delta F\equiv F_B-F_A$ in both equations.
As with the forward process, it is useful to distinguish between {\it typical}
realizations of the reverse process,
and the {\it dominant} realizations that contribute the most
to $\langle e^{-\beta W}\rangle_R$.

\begin{figure}[htbp]
  \centering
  \includegraphics[scale=0.60,angle=-90]{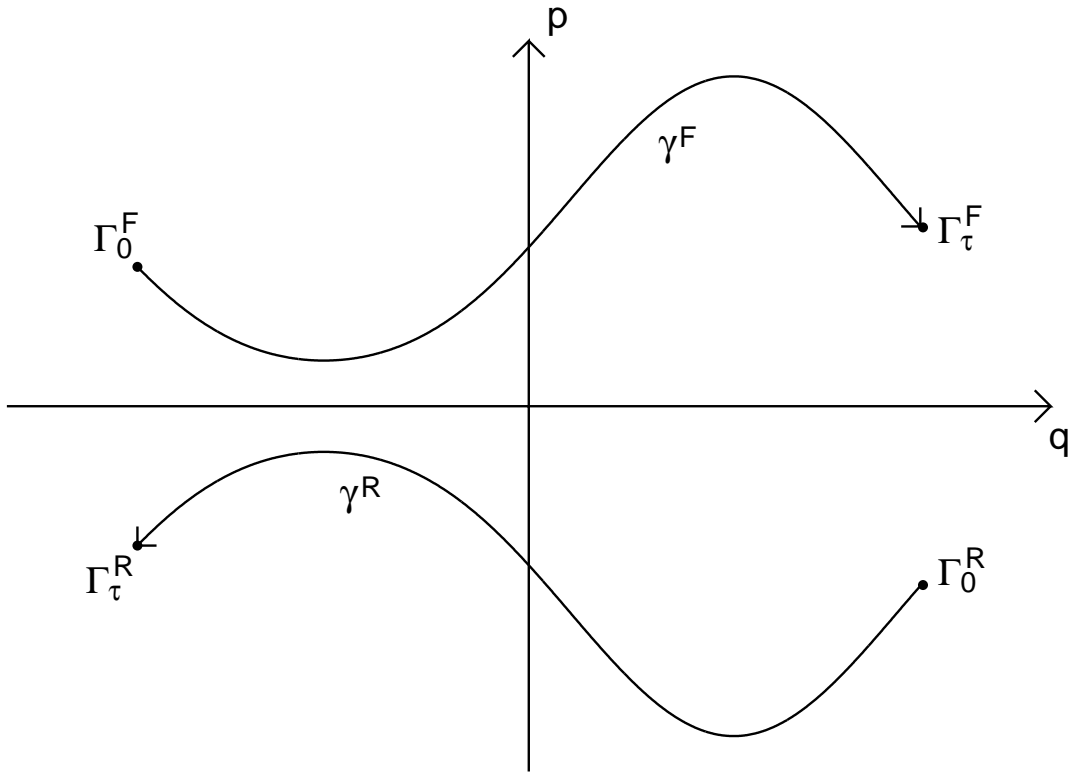}
  \caption{
A conjugate pair of trajectories.
The horizontal axis (${\bf q}$) represents the complete set of configurational coordinates
(e.g.\ particle positions), while the vertical axis (${\bf p}$) represents the set of
associated momenta.
$\Gamma$ denotes a point in this many-dimensional phase space;
and $\Gamma_t^{F/R}$ denotes a realization of the forward / reverse process, with
time running from $t=0$ to $t=\tau$.
The two trajectories are related by time-reversal:
$\Gamma_t^R = \Gamma_{\tau-t}^{F*}$, where
$({\bf q},{\bf p})^* \equiv ({\bf q},-{\bf p})$.
In the notation of Section \ref{sec:derivation}, 
the upper curve is the trajectory $\gamma^F$,
the lower curve is the trajectory $\gamma^R$.
}
  \label{fig:conjugatePair}
\end{figure}

Throughout this paper, the evolution of the system
is modeled as a Hamiltonian trajectory in phase space,
where the Hamiltonian is made time-dependent by externally varying the work parameter.
(See Section \ref{sec:discussion} for a brief discussion of other, e.g.\ stochastic, models.)
In the absence of magnetic fields~\footnote{
When magnetic fields are present, the conjugate pairing of trajectories
occurs if the reverse process is defined not only by Eq.~\ref{eq:for_rev_schedules},
but also by a change in the signs of these fields~\cite{Crooks99}.
The central result of this paper then remains valid.
For simplicity of presentation, however, I will restrict myself to the
time-reversal-invariant situation, Eq.~\ref{eq:tri}.
}
-- more precisely,
under the assumption of time-reversal invariance, Eq.\ref{eq:tri} --
forward and reverse realizations come in {\it conjugate pairs} related by time-reversal,
as illustrated in Fig.\ref{fig:conjugatePair}:
if a phase space trajectory $\Gamma_t^F$ represents a possible realization of the forward process
(a solution of Hamilton's equations when $\lambda$ is varied from $A$ to $B$), then its conjugate twin, 
\begin{equation}
\label{eq:conjPair}
\Gamma_t^R = \Gamma_{\tau-t}^{F*},
\end{equation}
represents a possible realization of the reverse process.
The asterisk denotes a reversal of momenta:
$({\bf q},{\bf p})^* = ({\bf q},-{\bf p})$.
The trajectory $\Gamma_t^R$ depicts the sequence of events that we would observe,
if we were to film the forward realization $\Gamma_t^F$ and then run the movie backward.

The central result of this paper, Eq.~\ref{eq:central} below, states that 
the {\it dominant} realizations of the forward process are
the conjugate twins of {\it typical} realizations of the reverse process, and vice versa.
Thus, the trajectories that contribute the most to $\langle e^{-\beta W}\rangle_F$,
are those during which the behavior of the system appears {\it as though we had
filmed a typical realization of the reverse process, and then run the movie backward}.
The existence of such a duality was anticipated by Ritort,
who observed that for large systems the work performed during a dominant realization
of one process is (minus) the work performed during a typical realization of the conjugate process;
see comments following Eq.~58 of Ref.~\cite{Ritort04}.

As an illustration of this result,
consider an ideal gas of (mutually non-interacting)
point particles inside a box closed off at one end by a piston,
and imagine that we act on the gas by pulling the piston outward.
In the context of Eq.\ref{eq:nwt0}, this system has recently been studied
by several groups~\cite{LuaGrosberg05,Sung05,Lua05,Bena05}.
Of particular relevance to the present paper is the analysis
of Lua and Grosberg~\cite{LuaGrosberg05}, who showed by explicit
calculation that, in the fast piston limit,
the dominant realizations are characterized by particles with initial
velocities sampled from deep within the tail of a Maxwellian distribution;
and for the reverse process, when the piston is pushed into the gas,
the dominant realizations are those for which there are no particle-piston collisions~\cite{Lua05}.
These conclusions are consistent with the discussion below.

\begin{figure}[htbp]
  \centering
  \includegraphics[scale=0.65]{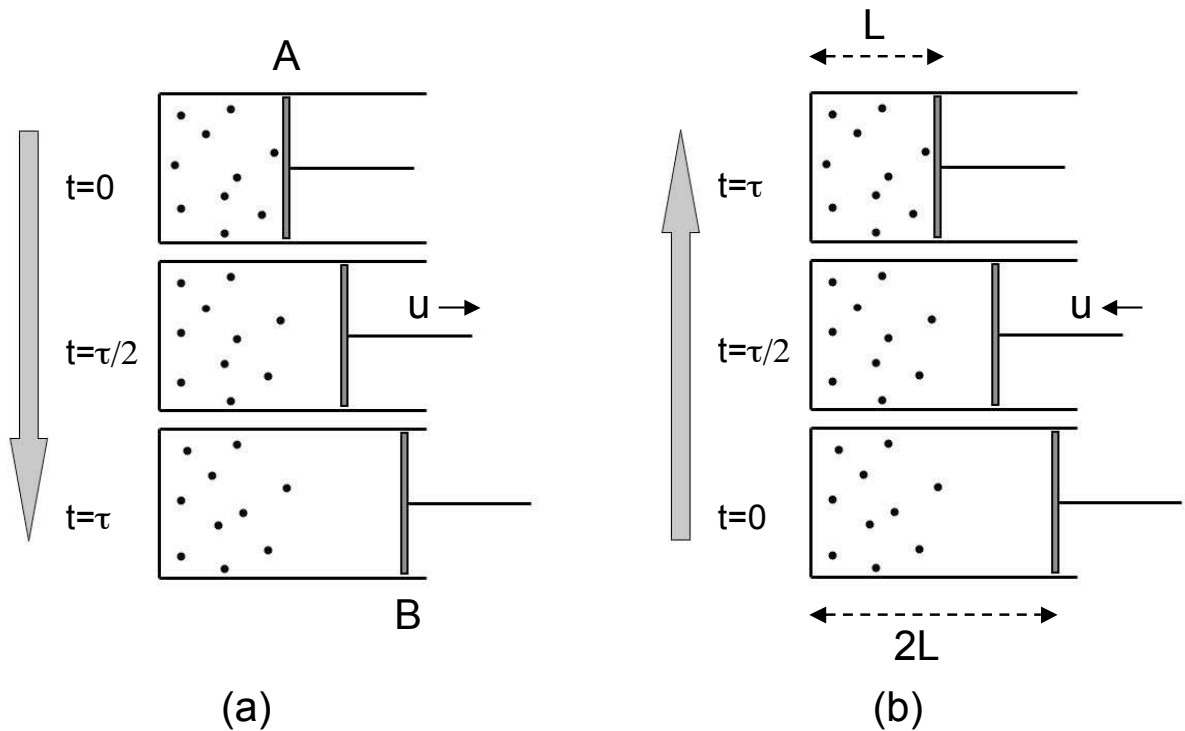}
  \caption{
(a) A typical realization of the forward process.
Due to the great speed of the piston (Eq.~\ref{eq:fastPiston}), most particles
remain in the left half the box for the duration of the process.
(b) The conjugate twin of the realization depicted in (a).
Almost all particles begin in the left half of the box,
and the piston then moves rapidly through a largely empty region.
(The vertical gray arrows specify the direction of increasing time.)
}
  \label{fig:piston_Ftyp_Rdom}
\end{figure}

Let $n_p \gg 1$ be the total number of particles, each of mass $m$.
Imagine that we begin with the piston at a location $A$, 
corresponding to a box of length $L$, and
we prepare the gas in canonical equilibrium at temperature $T$,
i.e.\ with particle velocities sampled independently from
a Maxwellian distribution.
Now we rapidly pull the piston outward,
from $A$ to $B$,
over a time $\tau$ and at constant speed $u=L/\tau$,
thus increasing the length of the box from $L$ to $2L$
(Fig.\ref{fig:piston_Ftyp_Rdom}a).
Since the volume of the box is doubled,
the free energy difference between the equilibrium states corresponding to
the piston locations $A$ and $B$ is
\begin{equation}
\Delta F = -n_p \beta^{-1} \ln 2.
\end{equation}
During a realization of this process, whenever a particle collides with the moving piston,
the particle suffers a change of kinetic energy,
$\delta K = -2m u ( v_x - u ) < 0$,
where $v_x$ is the component of the particle's velocity
parallel to the motion of the piston, prior to the collision.
The total work $W$ is the sum of such contributions.

\begin{figure}[htbp]
  \centering
  \includegraphics[scale=0.65]{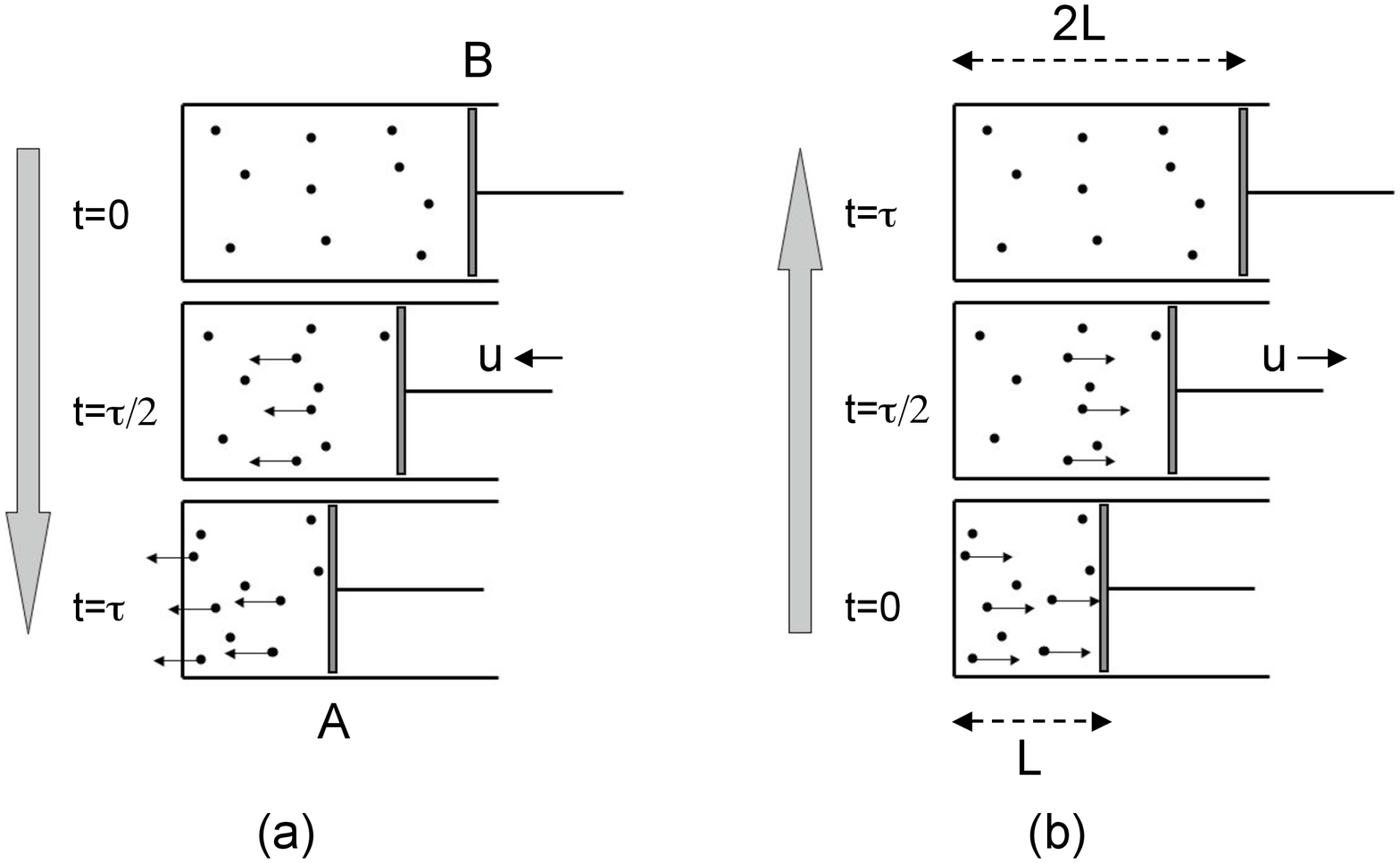}
  \caption{
(a) A typical realization of the reverse process.
As the piston moves rapidly into the gas, the particles with which it collides
gain large components of velocity ($\sim 2u$) along the direction of motion of the
piston.
The particles with the attached arrows are meant to represent these fast particles,
while the unadorned ones are characterized by thermal velocities.
(b) The conjugate twin of the realization depicted in (a).
Half the particles are initially moving at great speeds
($\sim 2u$) in the direction of the piston.
By the end of the process, after each of these has collided with the piston,
the velocity distribution of the entire gas is thermal.
}
  \label{fig:piston_Rtyp_Fdom}
\end{figure}

If the process described above is the {\it forward} process, then the
{\it reverse} process involves pushing the piston into the gas at speed $u$,
from $B$ to $A$ (Fig.\ref{fig:piston_Rtyp_Fdom}a),
starting from an initial state of thermal equilibrium.
Each particle-piston collision now produces a change
$\delta K = 2m u ( v_x + u ) > 0$
in the particle's kinetic energy.
In the context of the piston and gas example, I will use the terms
{\it expansion} and {\it compression} to denote the forward and reverse
processes, respectively.

Now suppose the piston speed is much greater than the thermal particle speed:
\begin{equation}
\label{eq:fastPiston}
u \gg v_{\rm th}=\sqrt{3/m\beta}.
\end{equation}
In this case, the particle density profile typically changes very little
during expansion (Fig.\ref{fig:piston_Ftyp_Rdom}a):
most particles remain in the left half of the box,
and few if any collide with the piston, consequently
\begin{equation}
W^F \approx 0.
\end{equation}
(The superscript indicates the forward process, i.e.\ expansion.)

During the compression process (Fig.\ref{fig:piston_Rtyp_Fdom}a),
the piston typically collides with about half of the gas particles 
-- roughly speaking, those initially located in right half of the box.
This generates a shock wave of particles
streaming leftward at approximately twice the piston speed.
At time $\tau$ the front of this wave reaches the wall at $x=0$,
just as the piston arrives at $A$.
Thus at the end of such a realization,
half the particles (those untouched by the piston)
are characterized by the original Maxwellian velocity distribution,
while the other half have velocities with $v_x \approx -2u$.
For such a realization,
\begin{equation}
W^R\approx \frac{n_p}{2}\cdot 2mu^2 = n_p m u^2 .
\end{equation}

If Fig.~\ref{fig:piston_Rtyp_Fdom}a illustrates the trajectory just described,
then Fig.\ref{fig:piston_Rtyp_Fdom}b illustrates
its conjugate twin.
Here the piston begins at $A$, with
half the particles streaming rightward at $v_x \approx +2u$.
As the piston moves from $A$ to $B$ at speed $u$,
each of these very fast particles collides once with the piston,
losing most of its kinetic energy.
At the moment the piston reaches $B$,
the container is uniformly filled with a gas characterized
by a Maxwellian velocity distribution at temperature $T$.
Needless to say, this ``anti-shock'' wave represents an exotic sequence of events!
When $n_p\gg 1$ and $u\gg v_{\rm th}$, the probability of sampling
an initial microstate for which half the particles have
$v_x \approx +2u$ is fantastically small.
However, according to the central result of this paper,
rare realizations of this sort are precisely the dominant
ones that contribute most to $\langle e^{-\beta W}\rangle_F$.

Similarly, the dominant realizations of the compression process
are the conjugate twins of typical realizations of the expansion process.
Thus to achieve convergence of $\langle e^{-\beta W}\rangle_R$, we must observe
realizations during which
the bulk of the gas happens to be localized in the left half of the box at $t=0$.
Then, as the piston moves rapidly from $B$ to $A$,
it sweeps through a largely empty region, as in
Fig.\ref{fig:piston_Ftyp_Rdom}b.
Again, this represents an unusual scenario:
it is unlikely that the randomly sampled initial conditions
are such that virtually all the particles are found in the left half of the box.

To summarize, while a typical realization of the expansion process is characterized
by no piston-particle collisions,
and a typical realization of the compression process is characterized by
$\sim n_p/2$ collisions (Figs.~\ref{fig:piston_Ftyp_Rdom}a, \ref{fig:piston_Rtyp_Fdom}a),
for {\it dominant} realizations it is the other way around
(Figs.~\ref{fig:piston_Rtyp_Fdom}b, \ref{fig:piston_Ftyp_Rdom}b).

\section{Derivation of central result}
\label{sec:derivation}

Consider a classical system described by a Hamiltonian $H(\Gamma;\lambda)$,
where $\Gamma =({\bf q},{\bf p})$ denotes a point in the system's phase space
(a {\it microstate}),
and $\lambda$ is an externally controlled parameter,
such as the piston location in the above example.
For every value of $\lambda$, assume $H$ is time-reversal invariant:
\begin{equation}
\label{eq:tri}
H(\Gamma^*;\lambda) = H(\Gamma;\lambda)
\quad,\quad
\Gamma^*\equiv({\bf q},-{\bf p}).
\end{equation}
This system can be prepared in an equilibrium state,
by placing it in weak thermal contact with a sufficiently large heat reservoir 
at temperature $T$, holding the parameter fixed at a value $\lambda$,
and then removing the reservoir after a sufficiently long relaxation time.
This generates a microstate $\Gamma_0$
that is a random sample from the Boltzmann-Gibbs distribution,
\begin{equation}
\label{eq:pBG}
p_\lambda(\Gamma_0) = \frac{1}{Z_\lambda} \exp[-\beta H(\Gamma_0;\lambda)],
\end{equation}
where $Z_\lambda = \int d\Gamma\,\exp[-\beta H(\Gamma;\lambda)]$ is the
partition function.
The free energy associated with this equilibrium state is
\begin{equation}
\label{eq:fdef}
F_\lambda = -\beta^{-1}\ln Z_\lambda.
\end{equation}
In the analysis and discussions below, the dependence
of $p_\lambda$, $Z_\lambda$, and $F_\lambda$ on temperature will be left implicit,
and equilibrium states will be identified by the parameter value $\lambda$.

To perform the forward process ($F$), we first prepare the system
in the equilibrium state $A$, then we remove the heat reservoir.
Then, from $t=0$ to $t=\tau$ we let the system evolve under Hamilton's equations as we vary $\lambda$
from $A$ to $B$ according to a pre-determined schedule,
$\lambda_t^F$.
Let $H_t^F(\Gamma)=H(\Gamma;\lambda_t^F)$ denote the time-dependent Hamiltonian obtained
when $\lambda$ is varied in this manner,
and let $\gamma^F = [\Gamma_t^F]_0^\tau$ denote a phase space trajectory evolving under
this Hamiltonian.
The notation indicates that the trajectory $\gamma^F$
passes through the set of points $\Gamma_t^F$, for $0\le t\le\tau$.
Such a trajectory describes the microscopic history of the system during a single
realization of the forward process.

By repeating this process infinitely many times, we generate a sequence of trajectories,
$\{\gamma_1^F,\gamma_2^F,\cdots\}$.
These can be viewed as random samples from a probability distribution
${\cal P}^F[\gamma^F]$,
defined on the set of all possible trajectories generated by $H_t^F$.
Because the dynamics are deterministic,
the probability of observing a given trajectory
is simply that of sampling its initial conditions
from a canonical distribution~\cite{Sun03}:
\begin{equation}
\label{eq:pa}
{\cal P}^F[\gamma^F] = 
p_A(\Gamma_0^F) =
\frac{1}{Z_A}\exp[-\beta H(\Gamma_0^F;A)].
\end{equation}
Since the system is thermally isolated (not in contact with a heat reservoir)
as $\lambda$ is varied from $A$ to $B$, the work performed on the system
is equal to the net change in its energy:
\begin{equation}
\label{eq:WFdef}
W^F[\gamma^F] = H(\Gamma_\tau^F;B) - H(\Gamma_0^F;A).
\end{equation}

Similar remarks and notation apply to the reverse process ($R$).
The system is prepared in equilibrium state $B$,
then $\lambda$ is varied from $B$ to $A$
(Eq.~\ref{eq:for_rev_schedules}),
generating a Hamiltonian trajectory $\gamma^R = [\Gamma_t^R]_0^\tau$,
with probability
\begin{equation}
\label{eq:pb}
{\cal P}^R[\gamma^R] = 
p_B(\Gamma_0^R) =
\frac{1}{Z_B}\exp[-\beta H(\Gamma_0^R;B)].
\end{equation}
The work performed is
\begin{equation}
\label{eq:WRdef}
W^R[\gamma^R] = H(\Gamma_\tau^R;A) - H(\Gamma_0^R;B).
\end{equation}

As mentioned in Section \ref{sec:summary},
every forward trajectory $\gamma^F$ has a conjugate twin $\gamma^R$
(Fig.~\ref{fig:conjugatePair}), that
is a solution of Hamilton's equations when the parameter is
varied according to the reverse schedule.
In the remainder of this paper, whenever $\gamma^F$ and $\gamma^R$
appear together (e.g.\ in the same equation or sentence)
it will be understood that these two trajectories
form such a conjugate pair.

From Eqs.\ref{eq:tri}, \ref{eq:WFdef} and \ref{eq:WRdef}, we get
\begin{equation}
\label{eq:wfr}
W^F[\gamma^F] = -W^R[\gamma^R] \quad;
\end{equation}
the work during a forward realization
is the opposite of that during its conjugate twin.
In what follows it will be convenient to deal with
{\it dissipated} work values
\begin{subequations}
\label{eq:defWd}
\begin{eqnarray}
W_d^F[\gamma^F] &\equiv& W^F[\gamma^F] - \Delta F \\
W_d^R[\gamma^R] &\equiv& W^R[\gamma^R] + \Delta F = - W_d^F[\gamma^F].
\end{eqnarray}
\end{subequations}
(The term ``dissipated work'' here means the amount by
which the work $W$ exceeds that which would have been
performed, had the process been carried out reversibly {\it and isothermally}.)
In terms of these quantities, Eq.~\ref{eq:both} becomes
\begin{equation}
\label{eq:both_diss}
\Bigl\langle e^{-\beta W_d}\Bigr\rangle_F =  1
\qquad,\qquad
\Bigl\langle e^{-\beta W_d}\Bigr\rangle_R = 1 .
\end{equation}

We now have the elements in place to investigate the nature of those realizations
that dominate the average
$\langle e^{-\beta W}\rangle_F$, or, equivalently,
$\langle e^{-\beta W_d}\rangle_F$.
Combining Eqs.\ref{eq:conjPair}, \ref{eq:tri}, \ref{eq:fdef}-\ref{eq:pb},
we obtain a simple relationship between the probability of observing a trajectory
$\gamma^F$ during the forward process,
and that of observing its twin $\gamma^R$ during the reverse:
\begin{equation}
\label{eq:ratio}
\frac{{\cal P}^F[\gamma^F]}{{\cal P}^R[\gamma^R]} =
\exp\Bigl( \beta W_d^F[\gamma^F] \Bigr) =
\exp\Bigl( -\beta W_d^R[\gamma^R] \Bigr).
\end{equation}
This result, originally obtained by Crooks
for stochastic, Markovian dynamics~\cite{Crooks98},
has here been derived in the context of Hamiltonian evolution.

(In Eqs.\ref{eq:pa} and \ref{eq:pb}, as in Ref.~\cite{Sun03},
the Liouville measure on initial conditions
in phase space has implicitly been used to define a measure on the space of
trajectories: the ``volume'' of trajectory space, $d\gamma^F$, associated with
a collection of forward trajectories,
is taken to be the phase space volume $d\Gamma_0^F$ occupied by their initial conditions.
Liouville's theorem then implies that the volume occupied by a
given set of forward trajectories is equal to that of
the conjugate set of reverse trajectories.
In this sense, the numerator and denominator in Eq.\ref{eq:ratio} are
defined with respect to the same measure on trajectory space.)

\begin{figure}[htbp]
  \centering
  \includegraphics[scale=0.60]{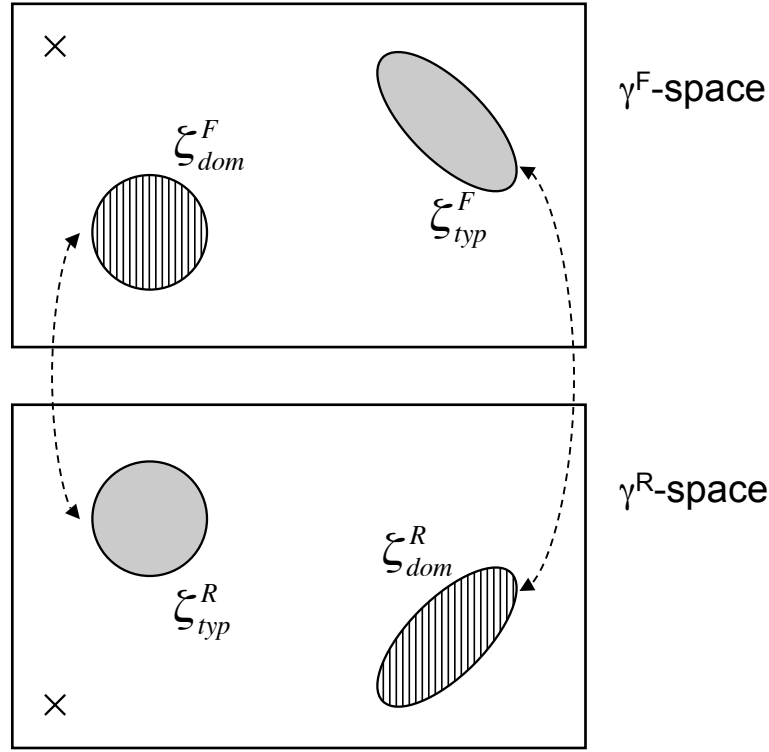}
  \caption{
The upper box represents the space of all
forward trajectories, the lower box the space of reverse trajectories,
and in this visual depiction conjugate pairing is indicated by
reflection about a horizontal line between the two boxes
(e.g.\ the two crosses represent a pair of conjugate twins).
The grey regions $\zeta_{\rm typ}^F$ and $\zeta_{\rm typ}^R$ denote
the peaks of the probability distributions ${\cal P}^F$ and ${\cal P}^R$,
while the vertically striped regions $\zeta_{\rm dom}^F$ and $\zeta_{\rm dom}^R$
are the peaks of the functions ${\cal Q}^F$ and ${\cal Q}^R$.
The dashed arrows indicate the conjugate pairing
that is the central result of this paper (Eq.~\ref{eq:central}).
}
  \label{fig:pathSpace}
\end{figure}

Let us now write $\langle e^{-\beta W_d}\rangle_F$ as an
integral over forward trajectories:
\begin{equation}
\label{eq:exp_F.q}
1 = \Bigl\langle e^{-\beta W_d}\Bigr\rangle_F = 
\int d\gamma^F\, {\cal P}^F[\gamma^F] \, e^{-\beta W_d^F[\gamma^F]} \equiv
\int d\gamma^F\, {\cal Q}^F[\gamma^F] ,
\end{equation}
where ${\cal Q}^F  = {\cal P}^Fe^{-\beta W_d^F}$.
Let $\zeta_{\rm typ}^F$ and $\zeta_{\rm dom}^F$ denote the regions
of trajectory space where ${\cal P}^F$ and ${\cal Q}^F$, respectively, are peaked,
as illustrated in Fig.\ref{fig:pathSpace}.
Thus, while $\zeta_{\rm typ}^F$ contains the {\it typical} forward realizations,
$\zeta_{\rm dom}^F$ contains the {\it dominant} ones, since
the greatest contribution in Eq.\ref{eq:exp_F.q}
comes from the peak region of ${\cal Q}^F$.

For the reverse process, we have
\begin{equation}
1 = \Bigl\langle e^{-\beta W_d}\Bigr\rangle_R = 
\int d\gamma^R {\cal P}^R[\gamma^R] \, e^{-\beta W_d^R[\gamma^R]} \equiv
\int d\gamma^R\, {\cal Q}^R[\gamma^R],
\end{equation}
and we define regions $\zeta_{\rm typ}^R$ and $\zeta_{\rm dom}^R$
where ${\cal P}^R$ and ${\cal Q}^R$ are peaked.

Combining Eq.\ref{eq:ratio} with
the definitions of ${\cal Q}^F$ and ${\cal Q}^R$, we get
\begin{subequations}
\label{eq:qFpR.qRpF}
\begin{eqnarray}
\label{eq:qFpR}
{\cal Q}^F[\gamma^F] &=& {\cal P}^R[\gamma^R] \\
\label{eq:qRpF}
{\cal Q}^R[\gamma^R] &=& {\cal P}^F[\gamma^F] .
\end{eqnarray}
\end{subequations}
Eq.~\ref{eq:qFpR} states that the function ${\cal Q}^F$
is the conjugate image of the distribution ${\cal P}^R$;
thus if we plot ${\cal P}^R$
in the lower box of Fig.~\ref{fig:pathSpace}, then its
mirror image in the upper box is ${\cal Q}^F$.
It follows that the trajectories in
$\zeta_{\rm dom}^F$  (the peak region of ${\cal Q}^F$)
are the conjugate twins of those in
$\zeta_{\rm typ}^R$ (the peak region of ${\cal P}^R$):
\begin{subequations}
\label{eq:central}
\begin{equation}
\label{eq:central1}
\zeta_{\rm dom}^F \leftrightarrow \zeta_{\rm typ}^R,
\end{equation}
where the symbol $\leftrightarrow$ indicates a correspondence through
conjugate pairing of trajectories.
This is illustrated by the pair of circles in Fig.~\ref{fig:pathSpace},
depicting the peak regions of ${\cal Q}^F$ and ${\cal P}^R$.
Similarly, Eq.~\ref{eq:qRpF} gives us
\begin{equation}
\label{eq:central2}
\zeta_{\rm dom}^R \leftrightarrow \zeta_{\rm typ}^F,
\end{equation}
\end{subequations}
as illustrated by the ellipses in Fig.~\ref{fig:pathSpace}.
Thus the trajectories typically observed during the reverse process
are the conjugate twins of those that dominate
$\langle e^{-\beta W}\rangle$ for the forward process (Eq.\ref{eq:central1}),
and vice-versa (Eq.\ref{eq:central2}).
This is the central result of this paper.

Let us consider for a moment the special case of a {\it cyclic} process, for which
the final value of the work parameter is the same as the initial value:
$\lambda_0^F = A = B = \lambda_\tau^F$.
In this situation $\Delta F = 0$, identically, and Eq.~\ref{eq:nwt0} becomes
\begin{equation}
\label{eq:bk}
\Bigl\langle e^{-\beta W}\Bigr\rangle_F = 1,
\end{equation}
a result originally derived
by Bochkov and Kuzovlev~\cite{BochkovKuzovlev}.
Under the additional assumption of a time-symmetric schedule,
$\lambda_t^F = \lambda_{\tau-t}^F$,
the forward and reverse processes are identical.
Thus,  for processes that are time-symmetric (and therefore also cyclic),
there is no distinction between ``forward'' and ``reverse'', and
Eq.~\ref{eq:central} is particularly easy to state:
the exponential average is dominated by the conjugate twins of typical realizations.

In the analysis leading to Eq.~\ref{eq:central}, it has implicitly been assumed that
the distributions ${\cal P}^F[\gamma^F]$ and ${\cal P}^R[\gamma^R]$
are sharply peaked,
i.e.\ that each process is characterized by
well-defined ``typical behavior''.
This assumption is often reasonable for systems with many degrees of freedom
(see e.g.\ Figs.~\ref{fig:piston_Ftyp_Rdom}a, \ref{fig:piston_Rtyp_Fdom}a).
It is useful, however, to formalize and generalize the discussion,
so as to avoid reliance on the notion of typicality.

Let $S^F$ denote an arbitrary set of forward trajectories, and define
\begin{equation}
\Omega^F\{S^F\} \equiv \int_{S^F} d\gamma^F {\cal P}^F[\gamma^F]
\qquad,\qquad
\Psi^F\{S^F\} \equiv
\int_{S^F} d\gamma^F {\cal Q}^F[\gamma^F] .
\end{equation}
$\Omega^F$ is the probability
of obtaining a trajectory in $S^F$,
when carrying out the forward process;
I will refer to this as the {\it statistical weight} of the set $S^F$
in the ensemble of forward trajectories.
In turn, $\Psi^F$ provides a measure of the {\it relative contribution} of $S^F$
to the average $\langle e^{-\beta W}\rangle_F$ (see Eq.~\ref{eq:exp_F.q}).
Note that $\Omega^F=\Psi^F=1$
when $S^F$ includes all forward trajectories.
Define analogous quantities
$\Omega^R$ and $\Psi^R$
for a set of reverse trajectories $S^R$.
Now take these two sets to be related by conjugate pairing: 
$S^F$ is an arbitrary region in the upper box of Fig.~\ref{fig:pathSpace},
and $S^R$ is its mirror image in the lower box.
Using Eq.\ref{eq:qFpR.qRpF} we then get
\begin{equation}
\label{eq:central_gen}
\Psi^F\{S^F\} = \Omega^R\{S^R\}.
\end{equation}
In words: the relative contribution of a set of forward trajectories
to $\langle e^{-\beta W}\rangle_F$, is equal to the statistical weight
of the conjugate set of reverse trajectories.
Of course, the converse is true as well:
$\Psi^R\{S^R\} = \Omega^F\{S^F\}$.

Eq.~\ref{eq:central} is a special case of Eq.\ref{eq:central_gen}:
if $\zeta_{\rm typ}^R$ contains 95\% of
the statistical weight of all reverse trajectories
(a reasonable definition of the peak region of ${\cal P}^R$),
then the conjugate set of forward trajectories
provides 95\% of the contribution to $\langle e^{-\beta W}\rangle_F$.

As an illustration of Eq.~\ref{eq:central_gen} using the piston-and-gas example,
consider the set ${\cal S}_n^F$ of all forward realization
for which there are exactly $n$ piston-particle collisions,
and the conjugate set ${\cal S}_n^R$ of reverse trajectories.
Then
\begin{equation}
\label{eq:central_gen.example}
\Psi^F\{S_n^F\} = \Omega^R\{S_n^R\},
\end{equation}
i.e.\ the relative contribution of $n$-collision realizations to
$\langle e^{-\beta W}\rangle_F$, is exactly the probability of observing
$n$ collisions when performing the reverse process, for any $n\ge 0$.
Since $\Omega^R\{S_n^R\}$ is peaked around $n\approx n_p/2$
(during compression we almost always observe roughly $n_p/2$ collisions,
Fig.~\ref{fig:piston_Rtyp_Fdom}a),
it follows that the greatest contribution to 
$\langle e^{-\beta W}\rangle_F$ comes from realizations of the expansion
process for which $n \approx n_p/2$
(Fig.~\ref{fig:piston_Rtyp_Fdom}b).

\section{Number of realizations needed for convergence}
\label{sec:numbers}

When using Eq.\ref{eq:estDF} to evaluate $\Delta F$,
how many realizations do we need to obtain a reasonable estimate?
While a precise answer depends on the desired accuracy,
a back-of-the-envelope estimate can be derived as follows.

Imagine that we repeatedly carry out the forward process,
either in a laboratory experiment or using numerical simulations.
In doing so we generate trajectories
$\{\gamma_1^F, \gamma_2^F, \cdots \}$ 
sampled from ${\cal P}^F[\gamma^F]$.
Since the most important contribution to $\langle e^{-\beta W}\rangle_F$
comes from the region $\zeta_{\rm dom}^F$,
we must sample this region to get a decent estimate of the average.
The probability that a single randomly sampled trajectory falls within $\zeta_{\rm dom}^F$ is
\begin{eqnarray}
P &=&
\int_{\zeta_{\rm dom}^F} d\gamma^F\, {\cal P}^F[\gamma^F] =
\int_{\zeta_{\rm typ}^R} d\gamma^R\, {\cal P}^R[\gamma^R] \, \exp\Bigl(-\beta W_d^R[\gamma^R]\Bigr) \\
&\sim& \exp\Bigl(-\beta \overline W_d^R\Bigr) \int_{\zeta_{\rm typ}^R} d\gamma^R\, {\cal P}^R[\gamma^R]
\sim \exp\Bigl(-\beta \overline W_d^R\Bigr).
\end{eqnarray}
Here $\overline W_d^R$ is the average work
dissipated when performing the reverse process.
On the first line we have used Eqs.\ref{eq:ratio} and \ref{eq:central1},
and on the second we have used the fact that the trajectories $\gamma^R\in\zeta_{\rm typ}^R$
constitute most of the probability distribution ${\cal P}^R$.
Thus an estimate of the number of realizations
needed to obtain a single trajectory in $\zeta_{\rm dom}^F$ is:
\begin{subequations}
\label{eq:Nc}
\begin{equation}
\label{eq:NcF}
N_c^F = P^{-1} \sim \exp\Bigl(\beta \overline W_d^R\Bigr).
\end{equation}
The same argument gives us the expected number of reverse realizations needed to 
obtain a trajectory in $\zeta_{\rm dom}^R$:
\begin{equation}
\label{eq:NcR}
N_c^R \sim \exp\Bigl(\beta \overline W_d^F\Bigr).
\end{equation}
\end{subequations}

\begin{figure}[htbp]
  \centering
  \includegraphics[scale=0.45,angle=-90]{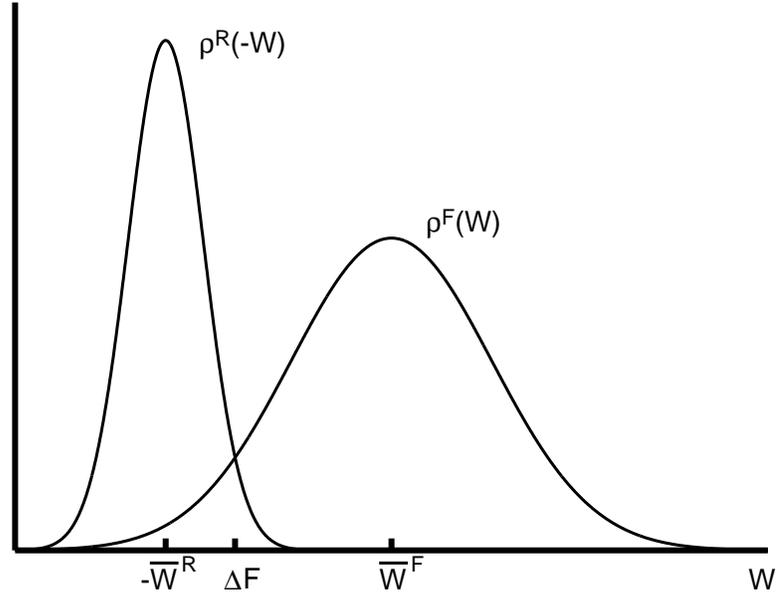}
  \caption{
Distributions of work values when $\overline W_d^F > \overline W_d^R$.
Since the two distributions cross at $W=\Delta F$,
$\rho^F$ is wider than $\rho^R$.
Thus, work values near $-\overline W^R$ are more frequently sampled
from $\rho^F(W)$,
than those near $\overline W^F$ are sampled from $\rho^R(-W)$.
}
  \label{fig:crossing}
\end{figure}

These results suggest that the number of realizations required for convergence
grows exponentially in the average dissipated work,
in agreement with the findings of Gore {\it et al.}~\cite{Gore03},
and therefore exponentially with system size (assuming dissipated work is an
extensive property), as concluded by Lua and Grosberg~\cite{LuaGrosberg05}.
Interestingly, however, it is the average amount of work dissipated during the {\it reverse} process
that determines the convergence of $\langle e^{-\beta W}\rangle$ for the {\it forward} process 
(Eq.\ref{eq:NcF}), and vice-versa (Eq.\ref{eq:NcR}).
This implies that, of the two processes,
the more dissipative one is the one for which
$\langle e^{-\beta W}\rangle$ converges more rapidly.
We can understand this counterintuitive conclusion
with the following plausibility argument.
Fig.~\ref{fig:crossing} depicts the work distributions $\rho^F(W)$ and $\rho^R(-W)$,
when $\overline W_d^F > \overline W_d^R$:
the mean of $\rho^F$ is displaced farther to the right of $\Delta F$
than the mean of $\rho^R$ is to the left of $\Delta F$, which
in turn suggests that $\rho^F$ is wider than $\rho^R$,
since the two distributions cross exactly at $W=\Delta F$~\cite{Collin05}.
Thus, as measured in standard deviations,
we must reach deeper into the tail of $\rho^R$
to sample the peak region of $\rho^F$,
than the other way around.
Hence $N_c^R > N_c^F$ when $\overline W_d^F > \overline W_d^R$.
This prediction agrees well with recent numerical simulations
of an asymmetric object dragged through a hard-disk gas~\cite{Cleuren05}.

The piston-and-gas example provides a nice illustration of Eq.~\ref{eq:Nc}.
From Section \ref{sec:summary} we have
$\overline W^F \approx 0$, $\overline W^R\approx n_pmu^2$, 
and $\Delta F = -n_p\beta^{-1}\ln 2$, thus
\begin{equation}
\label{eq:wd_gas}
\overline W_d^F \approx n_p\beta^{-1}\ln 2
\qquad,\qquad
\overline W_d^R \approx n_p(mu^2 - \beta^{-1}\ln 2).
\end{equation}
Hence $\overline W_d^R \gg \overline W_d^F$
in the fast piston limit (Eq.~\ref{eq:fastPiston});
rapid compression is much more dissipative than rapid expansion.

Let us now analyze the convergence of $\langle e^{-\beta W}\rangle_R$.
The dominant realizations are those for which
essentially all the gas particles begin in the left half of the box
(Fig.\ref{fig:piston_Ftyp_Rdom}b).
The probability of generating such an initial condition is
$(1/2)^{n_p}$, hence the number of realizations needed
to observe a single such event is 
\begin{equation}
N_c^R \sim 2^{n_p},
\end{equation}
which is equal to $\exp(\beta \overline W_d^F)$,
as predicted by Eq.\ref{eq:NcR}.

For the convergence of $\langle e^{-\beta W}\rangle_F$,
the dominant realizations are represented by Fig.\ref{fig:piston_Rtyp_Fdom}b:
initially, half the particles are characterized by a thermal velocity profile,
the other half with $v_x \approx 2u$.
The probability of generating such a microstate is roughly
\begin{equation}
\label{eq:roughEstimate}
\frac{n_p!}{(n_p/2)!(n_p/2)!} \cdot a^{n_p/2} .
\end{equation}
The first factor counts the number of ways of choosing which $n_p/2$ particles 
start with $v_x \approx 2u$,
and $a \equiv \exp[-\beta m (2u)^2/2]$ is an estimate of the probability for
a single particle to have a such a large initial $v_x$.
Using Stirling's approximation for the factorials, Eq.\ref{eq:roughEstimate}
reduces to
$2^{n_p} \cdot \exp(-n_p\beta mu^2)$.
Taking the reciprocal, we get
\begin{equation}
N_c^F \sim \exp[n_p(\beta mu^2 - \ln 2)],
\end{equation}
which is equal to $\exp(\beta \overline W_d^R)$,
as predicted by Eq.\ref{eq:NcF}.
Since the fast piston limit (Eq.~\ref{eq:fastPiston}) implies $\beta mu^2 \gg 1$,
it is legitimate to drop the $\ln 2$ term in this result, obtaining
$N_c^F \sim \exp(n_p\beta mu^2)$.
This is consistent with the calculations of Lua and Grosberg,
who obtain $N_c^F \sim \exp(\beta m L^2/\tau^2)$
when $n_p=1$ (see section 4 of Ref.~\cite{LuaGrosberg05}). 

These results verify that the more dissipative process
(compression) is the one for which the exponential average converges
more rapidly.
We can understand why this is the case, without performing explicit quantitative estimates.
For the compression process, as mentioned, a dominant realization begins with
all the particles in the left half of the box,
thus $n_p$ particles must simultaneously satisfy a condition 
that is not so unusual for any given particle.
For the expansion process, on the other hand,
half the particles must begin with $v_x \approx 2u$,
therefore $n_p/2$ particles must simultaneously satisfy a condition that
is very unusual for even a single particle (since $u\gg v_{\rm th}$).
The latter situation is the less likely by far.
Of course, when $n_p \gg 1$, both $N_c^F$ and $N_c^R$ are
extremely large~\cite{Gross.condmat0509648}.

In practice, the convergence difficulties associated with Eq.~\ref{eq:nwt0}
are mitigated somewhat if we have data
for both the forward and the reverse processes.
In that case, Bennett's {\it acceptance ratio method}~\cite{Bennett76,Crooks00,Shirts03,Collin05}
converges faster than a direct exponential average
of either the forward or reverse work values.
It would be useful to derive a simple estimate,
analogous to Eq.~\ref{eq:Nc},
of the number of realizations required for this method
to converge, and to develop heuristic insight regarding the realizations
that make the most important contribution to the acceptance ratio estimate of
$\Delta F$.

Finally, recall that a measure of the difference between two 
normalized distributions,
$f_1$ and $f_0$, is given by the {\it relative entropy}~\cite{coverThomas}
\begin{equation}
\label{eq:kldef}
D[f_1 \vert f_0] = \int f_1 \ln \frac{f_1}{f_0} \ge 0,
\end{equation}
where the integral is over the space of variables on which 
$f_0$ and $f_1$ are defined.
Applying this definition to the forward and reverse distributions of trajectories,
and identifying $\gamma^F$ with its twin $\gamma^R$,
we get
\begin{subequations}
\label{eq:relativeEntropy}
\begin{equation}
\label{eq:DFR}
D[{\cal P}^F \vert {\cal P}^R] = 
\int d\gamma^F\,{\cal P}^F[\gamma^F] \ln \frac{{\cal P}^F[\gamma^F]}{{\cal P}^R[\gamma^R]} =
\beta \,\overline W_d^F,
\end{equation}
using Eq.\ref{eq:ratio},
and similarly
\begin{equation}
\label{eq:DRF}
D[{\cal P}^R \vert {\cal P}^F] = 
\beta \,\overline W_d^R.
\end{equation}
\end{subequations}
Analogous identities have been derived
for the steady states of Markov chains~\cite{gaspard04},
and for the work distributions arising
in the context of free energy estimation~\cite{wuKofke};
and the physical significance of relative entropy for equilibrium and
nonequilibrium fluctuations has recently been discussed in Ref.~\cite{Qian01}.
Eq.~\ref{eq:relativeEntropy} suggests
that there might be a natural information-theoretic
interpretation of Eq.~\ref{eq:Nc}.

\section{Free energy perturbation}
\label{sec:fep}

The {\it perturbation method} for estimating free energy differences is based on
the identity
\begin{equation}
\label{eq:fep}
\Bigl\langle e^{-\beta\Delta H} \Bigr\rangle_A = e^{-\beta\Delta F},
\end{equation}
where $\Delta H(\Gamma) \equiv H(\Gamma;B) - H(\Gamma;A)$,
and $\langle\cdots\rangle_A$ denotes an average over microstates
$\Gamma$ sampled from the canonical ensemble $A$~\cite{zwanzig,frenkelSmit}.
[In this section, I explicitly assume that the Hamiltonians
$H(A)$ and $H(B)$ are finite-valued throughout phase space.
This precludes situations such as the piston-and-gas
example, particles with perfectly hard cores, etc.]
The convergence problems that arise with the nonequilibrium work theorem
also plague the free energy perturbation identity~\cite{ZuckermanWoolf04}.
In the case of Eq.\ref{eq:fep}, we can frame this issue by considering
the {\it typical} microstates sampled from the equilibrium ensemble $A$,
and the rare but {\it dominant} microstates that contribute most to the
average of $e^{-\beta\Delta H}$.
In what follows I discuss how the ideas developed in earlier sections
of this paper apply to Eq.~\ref{eq:fep}.

The left side of Eq.~\ref{eq:fep} is most naturally viewed as an equilibrium
average, as described above.
An alternative perspective, however, treats Eq.~\ref{eq:fep} as a special
case of Eq.~\ref{eq:nwt0}, obtained in the ``instantaneous switching'' limit,
$\tau\rightarrow 0$, in which
$\lambda$ is changed suddenly from $A$ to $B$~\cite{Jarzynski97}.
The system then has no opportunity to evolve during the process,
hence a realization is described not by a trajectory,
but rather by a single microstate (sampled from $A$).
The work $W$ is given by $\Delta H$, evaluated at this microstate.
If we view Eq.\ref{eq:fep} as pertaining to a forward perturbation
($A\rightarrow B$),
then the reverse perturbation involves sampling
from equilibrium state $B$:
\begin{equation}
\label{eq:fepBA}
\Bigl\langle e^{+\beta\Delta H} \Bigr\rangle_B = e^{+\beta\Delta F},
\end{equation}
where $\Delta H$ and $\Delta F$ are defined
identically in Eqs.\ref{eq:fep} and \ref{eq:fepBA}.

By analogy with Eq.\ref{eq:exp_F.q} we can rewrite Eq.\ref{eq:fep} as follows:
\begin{equation}
1 =
\Bigl\langle e^{-\beta W_d^A}\Bigr\rangle_A = 
\int d\Gamma\, p_A(\Gamma) \, e^{-\beta W_d^A(\Gamma)} \equiv
\int d\Gamma\, q_A(\Gamma) ,
\end{equation}
where $p_A$ is the equilibrium distribution for state $A$
(Eq.\ref{eq:pBG}), and $W_d^A \equiv \Delta H - \Delta F$.
Similarly rewriting Eq.\ref{eq:fepBA} (with $W_d^B \equiv -\Delta H + \Delta F$),
let us now define $\xi_{\rm typ}^A$, $\xi_{\rm dom}^A$, $\xi_{\rm typ}^B$, and $\xi_{\rm dom}^B$
as the regions of phase space where
$p_A$, $q_A$, $p_B$, and $q_B$, respectively, are peaked.
These contain the typical and dominant microstates for the
two perturbations.
Eq.~\ref{eq:central} now becomes
\begin{equation}
\label{eq:xi}
\xi_{\rm dom}^A = \xi_{\rm typ}^B
\qquad,\qquad
\xi_{\rm dom}^B = \xi_{\rm typ}^A.
\end{equation}
Thus, when implementing the free energy perturbation method,
by sampling from one distribution (say, $A$),
the collection of sampled microstates must be large enough to
include a reasonable number that are typical of the other distribution ($B$),
otherwise we will not achieve convergence of the exponential average.
If there is very little overlap between the two equilibrium distributions
in phase space, then the number of samples required to satisfy this
condition is prohibitively large~\cite{frenkelSmit}.

Lower bounds $N_c^A$ and $N_c^B$ on the required numbers of realizations
are given by
analogues of Eq.~\ref{eq:Nc}:
\begin{subequations}
\label{eq:nanb}
\begin{eqnarray}
\ln N_c^A &\sim&
\beta \overline W_d^B
= D[p_B\vert p_A] \ge 0 \\
\ln N_c^B &\sim&
\beta \overline W_d^A
= D[p_A\vert p_B] \ge 0 ,
\end{eqnarray}
\end{subequations}
where the overbars now denote canonical averages with respect
to the ensembles $A$ and $B$.
Of the two perturbations, the one with larger $\overline W_d$
requires fewer samples for convergence of the exponential average.

These results are illustrated by Widom's particle insertion
method for computing a chemical potential~\cite{Widom63,frenkelSmit}.
Imagine a fluid of $N+1$ particles,
and suppose that $H(A)$ describes the situation in which
$N$ of the particles interact with one another through a pairwise potential,
while the remaining, ``tagged'' particle is uncoupled from the rest;
and $H(B)$ describes the situation in which all $N+1$ particles are
mutually coupled via the pairwise potential.
The free energy difference $\Delta F = F_B -F_A$
is the excess chemical potential of the fluid,
provided $N$ is large enough to recover bulk properties.
In principle,
we can estimate $\Delta F$ either by using the forward perturbation,
$A\rightarrow B$ (particle insertion),
or with the reverse perturbation, $B\rightarrow A$ (deletion).

In a dense fluid, particle insertion usually generates a very large
value of $\Delta H$,
while for deletion $\Delta H$ is typically modest.
Thus $\overline W_d^A > \overline W_d^B$, and
Eq.\ref{eq:nanb} predicts that the insertion method converges more rapidly than
the deletion method, as indeed observed empirically~\cite{frenkelSmit}.
To understand this in terms of typical and dominant microstates,
note that when sampling from ensemble $B$, the (interacting) tagged particle
typically occupies its own small volume within the fluid,
from which the remaining particles are excluded.
Thus, by Eq.~\ref{eq:xi}, to achieve convergence in Eq.\ref{eq:fep}
we must sample sufficiently from ensemble $A$ to obtain 
microstates in which the {\it non-interacting}, tagged particle happens to sit
inside a cavity created by the spontaneous fluctuations of the remaining
$N$-particle fluid.
This condition carries an entropic cost roughly
equal to the free energy of forming a suitably large cavity.

Conversely, when sampling from ensemble $A$
the (non-interacting) tagged particle is typically found
within the repulsive core of one of the other particles,
hence to succeed with the particle {\it deletion} method,
we must generate such microstates when sampling from ensemble $B$.
This carries an enthalpic (energetic) cost,
determined by the strength of the repulsive core of the pairwise potential.
Because the repulsive cores are generally described by very steep potentials,
this enthalpic penalty is much larger than the entropic penalty described above:
thermal fluctuations are much more likely to generate a cavity large enough
to accommodate a new particle,
than they are to squeeze two particles into a volume meant only for one.
Hence the dominant realizations of the forward perturbation (insertion),
are much less rare than those of the reverse perturbation (deletion).

\section{Discussion}
\label{sec:discussion}

The central result of this paper, Eq.~\ref{eq:central}, is a duality that relates
the dominant realizations of a given process to typical realizations of the conjugate process.
In a nutshell, it states that to achieve convergence in Eq.~\ref{eq:nwt0} 
we must observe realizations during which the system appears as though
it is evolving backward in time.
I will now sketch an interpretation of this result
similar to the discussion of {\it causal} and {\it anti-causal}
response theory found in Ref.~\cite{EvansSearles96}.
Following that, I will briefly discuss the validity of Eq.~\ref{eq:central}
in situations involving non-Hamiltonian (including stochastic) equations
of motion.

Let us picture the ensemble of forward trajectories,
$\{\gamma_1^F,\gamma_2^F,\cdots\}$,
as a swarm of points evolving independently in phase space
(from $t=0$ to $t=\tau$),
and let $p^F(\Gamma,t) = \langle \delta (\Gamma - \Gamma_t^F) \rangle$
denote the corresponding time-dependent phase space distribution.
This distribution obeys the Liouville equation,
\begin{equation}
\label{eq:liouville}
\frac{\partial p^F}{\partial t} =
\frac{\partial H_t^F}{\partial{\bf q}}\cdot\frac{\partial p^F}{\partial{\bf p}} -
\frac{\partial H_t^F}{\partial{\bf p}}\cdot\frac{\partial p^F}{\partial{\bf q}},
\end{equation}
where $H_t^F = H(\Gamma;\lambda_t^F)$.
The assumption of initial equilibrium
is a boundary condition imposed at $t=0$:
\begin{equation}
\label{eq:bc0}
p^F(\Gamma,0)
= \frac{1}{Z_A} \exp[-\beta H(\Gamma;A)].
\end{equation}
Now consider a distribution $q^F(\Gamma,t)$ evolving
under the same dynamics, Eq.\ref{eq:liouville}, but satisfying
a boundary condition at $t=\tau$ (rather than at $t=0$):
\begin{equation}
\label{eq:bctau}
q^F(\Gamma,\tau) 
= \frac{1}{Z_B} \exp[-\beta H(\Gamma;B)].
\end{equation}
This distribution describes an ensemble that ends, rather than
begins, in a state of thermal equilibrium.
In the language of Ref.~\cite{EvansSearles96}, $p^F$ corresponds to a {\it causal}
ensemble of trajectories, determined by initial conditions,
while $q^F$ is {\it anti-causal}, determined by final conditions.
The central result of this paper can now be restated as follows:
while the typical {\it causal} trajectories are the ones we ordinarily observe,
the typical {\it anti-causal} trajectories are the ones that
dominate the exponential average.
This follows from the simple observation that the anti-causal ensemble of forward
trajectories is just the conjugate image of the causal ensemble of reverse trajectories.

In the context of linear response theory, Evans and Searles~\cite{EvansSearles96}
have shown that anti-causal ensembles give rise to Green-Kubo ``anti-transport''
coefficients.
In an earlier theoretical study of dilute gases,
Cohen and Berlin~\cite{CohenBerlin60} derived an anti-causal version of
the Boltzmann equation, by applying the assumption of molecular chaos to
{\it future} rather than {\it past} pair distribution functions.
In both papers the anti-causal behavior is associated with violations of
the second law of thermodynamics:
the Green-Kubo coefficients of Ref.~\cite{EvansSearles96} have the ``wrong'' signs,
and the Boltzmann-like equation of Ref.~\cite{CohenBerlin60} obeys
an anti-$H$ theorem.
The situation is similar here: anti-causal ensembles
are associated with negative average values of dissipated work;
the second law of thermodynamics, by contrast, asserts that
irreversible processes are accompanied by positive average dissipated work.

Eq.~\ref{eq:ratio}
provides an amusing connection between the conjugate pairing of trajectories
(microscopic reversibility)
and the second law of thermodynamics (macroscopic irreversibility),
illustrated in the piston-and-gas context by the following thought experiment.
Imagine that we are shown a movie in which we see
the microscopic evolution of the gas as the piston moves outward from $A$ to $B$,
and we are asked
to guess whether this movie depicts an actual realization of the
expansion process,
or whether, instead, a realization of the compression process was filmed,
and now that movie is being run backward.
I will refer to this as ``guessing the direction of time''.
To analyze this situation quantitatively,
let $\gamma^F$ specify the microscopic evolution we observe when watching this
movie, and $\gamma^R$ its conjugate twin.
Then the task of guessing the direction of time is an exercise
in statistical inference, in which we compare the likelihoods of two hypotheses.
Specifically, we ask whether it is more likely
that we obtain $\gamma^F$ when performing the forward process,
or $\gamma^R$ when performing the reverse process.
By Eq.\ref{eq:ratio}, the ratio of these likelihoods is $\exp(\beta W_d^F)$.
Hence if $W_d^F > 0$, we opt for the first hypothesis,
namely that the piston was indeed withdrawn from $A$ to $B$;
whereas if $W_d^F< 0$ (equivalently $W_d^R>0$) then we guess that the piston was pushed into
the gas, and we are seeing a movie of that process in time-reversed order.
Thus when asked to guess the arrow of time,
we optimize our answer simply by insisting that the sign of the dissipated work be positive,
in agreement with the second law.

It is natural to use Hamilton's equations to describe a thermally isolated
classical system, as in this paper.
If the system is in contact with a heat reservoir, however,
then there are various ways to model its evolution.
The most intuitively natural is to treat the combined system
{\it and reservoir} as a very large, Hamiltonian system.
Alternatively, one can describe the evolution of the system itself using
stochastic equations of motion, such as the Metropolis Monte Carlo
algorithm, or Langevin dynamics.
Yet another approach involves deterministic but non-Hamiltonian equations
of motion,
such as Gaussian thermostats,
designed to mock up the presence of a heat reservoir.
Eq.\ref{eq:nwt0} has been derived for all these cases,
so it is natural to ask whether the central result of the present paper,
Eq.~\ref{eq:central},
also holds for these various schemes.

Since the heart of the argument in Section~\ref{sec:derivation} follows from
Eq.~\ref{eq:ratio},
two conditions are sufficient for the central results of this paper
to hold for any given one of the above-mentioned schemes.
First, there must exist a conjugate pairing of forward and reverse trajectories.
Second, the probability of observing a particular trajectory $\gamma^F$ during
the forward process, and that of observing its twin $\gamma^R$ during the
reverse process, must satisfy Eq.~\ref{eq:ratio}.
These conditions have been verified explicitly when the system
evolves under a discrete-time Monte Carlo scheme satisfying
detailed balance (such as the Metropolis algorithm)~\cite{Crooks98} or
under Langevin dynamics~\cite{Jarzynski98,Kurchan05,ChernyakChertkovJarzynski06}.
When the evolution of the system is modeled with isokinetic Gaussian equations
of motion, the validity of Eq.~\ref{eq:ratio} follows from
the analysis of Ref.~\cite{Evans03}.
Thus Eq.~\ref{eq:central} of the present paper applies when these schemes
are used to model the evolution of the system.

When the system and reservoir are treated together as a very large,
Hamiltonian system,
then Eq.~\ref{eq:ratio} can be derived by repeating
the steps of Section~\ref{sec:derivation},
but working in the {\it full} phase space containing all the interacting
degrees of freedom,
and then projecting out the reservoir degrees of freedom.
The analysis becomes slightly complicated if the coupling between
the system and reservoir is not negligible,
but this technical issue is handled much as in Ref.~\cite{Jarzynski04}
(see, however, Refs.~\cite{CohenMauzerall04,CohenMauzerall05}),
and the details will not be presented here.

Finally, the nonequilibrium work theorem is just one of
a number of  (mostly recent) predictions concerning the statistical mechanics of
systems far from thermal equilibrium.
Others include the Kawasaki identity~\cite{YamadaKawasaki67}
and its generalization by Morriss and Evans~\cite{MorrissEvans85}, the {\it fluctuation theorem}~\cite{EvansCohenMorriss93,EvansSearles94,GallavottiCohen95,Kurchan98,LebowitzSpohn99,Maes99,EvansSearles02,Wang-etal02},
Hatano and Sasa's equality for transitions between nonequilibrium steady
states~\cite{Hatano99,HatanoSasa01,Trepagnier04},
and Adib's microcanonical version of Eq.~\ref{eq:nwt0}~\cite{Adib.PRE.05}.
It remains to be investigated whether the analysis of the present paper
is valid (and relevant) in the context of these other, closely related results.

\vskip .2in

It is a pleasure to acknowledge useful correspondence with
Alexander Grosberg and Rhonald Lua.
This work was supported by the United States Department of Energy,
under contract W-7405-ENG-36.

\end{document}